\documentclass[10pt,journal]{IEEEtran}
\usepackage{setspace}
\usepackage{cite}
\usepackage{hhline}
\usepackage{amsmath,amssymb,amsfonts,bm}
\usepackage{dsfont}
\usepackage{amsthm}
\usepackage{algorithmic}
\usepackage{graphicx}
\usepackage{multirow}
\usepackage{float}
\usepackage{mleftright}
\usepackage{stfloats}
\usepackage{textcomp}
\usepackage{verbatim}
\usepackage[outdir=./]{epstopdf}
\usepackage{xcolor}
\usepackage{subcaption}
\usepackage[ruled,vlined]{algorithm2e}

\newcommand\undermat[2]{%
  \makebox[0pt][l]{$\smash{\underbrace{\phantom{%
    \begin{matrix}#2\end{matrix}}}_{\text{$#1$}}}$}#2}

\def\BibTeX{{\rm B\kern-.05em{\sc i\kern-.025em b}\kern-.08em
    T\kern-.1667em\lower.7ex\hbox{E}\kern-.125emX}}

\title{Advanced Channel Coding Designs for \\ Index-Modulated Fluid Antenna Systems}

\author{$\text{Elio Faddoul}$, \textit{Graduate Student Member, IEEE}, $\text{Ghassan M. Kraidy}$, \textit{Senior Member, IEEE},\\ $\text{Constantinos Psomas}$, \textit{Senior Member, IEEE}, and $\text{Ioannis Krikidis}$, \textit{Fellow, IEEE}\vspace*{3.75mm} \thanks{This work was presented in part at the IEEE Global Communications Conference, Kuala Lumpur, Malaysia, December 2023 \cite{Faddoul2023corr}.} \thanks{Elio Faddoul, Constantinos Psomas, and Ioannis Krikidis are with the Department of Electrical and Computer Engineering, University of
Cyprus, Nicosia, Cyprus (e-mail: \{efaddo01, psomas, krikidis\}@ucy.ac.cy). \\ Ghassan M. Kraidy is with the Department of Electronic Systems, Norwegian University of Science and Technology, Gj{\o}vik, Norway (e-mail: {ghassan.kraidy@ntnu.no}).}}
\begin{document}
\maketitle
\begin{abstract}
The concept of fluid antennas (FAs) has emerged as a promising solution to enhance the spectral efficiency of wireless networks, achieved by introducing additional degrees of freedom, including reconfigurability and flexibility. In this paper, we investigate the use of index-modulated (IM) transmissions within the framework of FA systems, where an FA position is activated during each transmission interval. This approach is motivated by the common characteristics exhibited by FAs and IM transmissions, which entails the use of a single radio-frequency chain. From this perspective, we derive a closed-form expression for the bit error rate of IM-FAs considering spatial correlation, demonstrating superior performance compared to conventional IM systems. To enhance the performance of IM-FAs in correlated conditions, channel coding techniques are applied. We first analyze a set partition coding (SPC) scheme for IM-FAs to spatially separate the FA ports, and provide a tight performance bound over correlated channels. Furthermore, the spatial SPC scheme is extended to turbo-coded modulation where the performance is analyzed for low and high signal-to-noise ratios. Our results reveal that through the implementation of channel coding techniques designed for FAs and IM transmission, the performance of coded IM-FAs exhibits notable enhancements, particularly in high correlation scenarios.
\end{abstract}


\section{Introduction}
During the past decades, various technology trends have been driving wireless communications forward. In particular, the field of antenna and radio frequency (RF) technologies has witnessed remarkable advancements to meet the requirements of next-generation communication systems \cite{mietzner2009multiple}. Among these developments, the utilization of multiple antenna technologies has garnered substantial research attention. Multiple-input multiple-output (MIMO) systems, in particular, are a pivotal wireless communication technology known for delivering notable diversity and multiplexing gains \cite{zhang2020prospective}. However, it is noteworthy that these antennas are usually constructed using fixed metallic structures, and are tailored to specific network requirements. Furthermore, the prevailing designs are susceptible to physical deployment limitations, including the imperative constraint of maintaining an inter-antenna separation of at least half a carrier's wavelength.

To circumvent the physical limits exhibited by conventional antennas, metamaterials have emerged as a promising technology whereby their potential lies in enabling the development of compact (small-sized) antennas \cite{ziolkowski2006metamaterial}. More recently, the authors in \cite{wong2020fluid} demonstrated that significant diversity gains can be achieved to reduce outage probabilities within confined spatial constraints, provided that antenna positions can be dynamically adjusted along fixed locations, otherwise known as ports. This concept is coined as fluid antenna (FA), which represents any software-controllable radiating fluidic structure that can alter its shape and position to reconfigure several parameters such as the operating frequency, polarization, and radiation patterns \cite{shojaeifard2022mimo,huang2021liquid}. Indeed, FA technologies fall under the wider concept of movable antennas, which may include designs that involve no fluidic materials to emulate the agility, such as mechanically-adjusted antennas using stepper motors \cite{ma2023mimo} or pixel antennas \cite{rodrigo2014frequency}. In this regard, several studies have focused on studying the performance of FAs from a communication theory perspective for point-to-point single-user systems \cite{wong2020fluid, wong2020performance, mukherjee2022level, psomas2023diversity}. For instance, initial works show that the outage probability of FA systems decreases as the number of ports increases, and can outperform maximal ratio combining for a large number of ports \cite{wong2020fluid}. The performance of FA systems is also analyzed in \cite{wong2020performance} in terms of the ergodic capacity under correlated Rayleigh fading channels, where closed-form expressions of the capacity lower bound are provided. More recently, closed-form analytical expressions of the average level crossing rate were derived for such FA systems \cite{mukherjee2022level}. The design of space-time coded modulations that exploit the FA’s sequential operation with space-time rotations and code diversity was analyzed as well in \cite{psomas2023diversity} for various FA architectures. Besides, the works in \cite{wong2021fluidmulti, wong2022fast, skouroumounis2022fluid} evaluate the performance of FAs for multi-user systems.

The performance of FA systems is highly influenced by the spatial correlation model employed to capture the interdependence among FA ports. Specifically, initial works adopt a generalized correlation model by treating the first FA port as a reference to capture the strong spatial correlation over the different FA ports \cite{wong2020fluid, mukherjee2022level, psomas2023diversity}. Although this model constitutes a fundamental initial measure towards acquiring knowledge of the FA technology, it does not accurately capture the correlation between the FA ports. In light of this, the authors in \cite{wong2022closed} derived a closed-form expression for spatial correlation parameters in the context of a point-to-point FA system. Furthermore, the work in \cite{khammassi2023new} derived an eigenvalue-based model that offers an accurate but complex approximation of spatial correlations among FA ports given by Jakes’ model. The authors in \cite{ghadi2023copula} recently introduced a copula-based framework for characterizing the joint multivariate distribution of correlated fading channels within the FA system which can describe spatial correlations among FA ports.

Moreover, as wireless communication systems continue to evolve, an imperative need arises to enhance both spectral and energy efficiencies. To address these challenges, the novel paradigm of index-modulated (IM) transmissions was introduced as a viable solution \cite{mao2018novel}. In IM systems, only a specific fraction of indexed resource entities are activated for data transmission, e.g., subcarriers (in orthogonal frequency division multiplexing) \cite{bacsar2013orthogonal}, antennas (in MIMO) \cite{jeganathan2009space}, or time slots. In this way, additional information bits are implicitly conveyed through patterns of active indices. Consequently, from a transceiver standpoint, an IM-aided system eliminates the need for precise time synchronization among antennas while consuming substantially less energy \cite{jeganathan2008spatial}. The concept of directional modulation has also been studied recently in \cite{wei2020secure}, with the main difference from IM transmissions being that modulation occurs at the antenna level rather than at the baseband, offering a new approach to physical layer security. In recent years, substantial efforts have been dedicated to incorporate practical coding strategies into space-domain IM transmissions \cite{jeganathan2009space, koca2012bit, mesleh2010trellis, basar2011new, feng2018nonbinary, dai2021polar, liu2022maximum}. For instance, the authors in \cite{jeganathan2009space} present a bit-interleaved coded modulation (BICM) system using iterative decoding for space-shift keying, where both conventional and turbo-coded modulation are employed. The principle of BICM is later extended in \cite{koca2012bit} to encompass spatial modulation (SM), enhancing error protection for both antenna and symbol bits. Furthermore, to improve the performance in correlated channels, the authors in \cite{mesleh2010trellis} proposed a trellis-coded SM scheme for spatially correlated MIMO systems in which the antenna index bits were processed by a convolutional encoder. Nonetheless, this method was confined to a predetermined code structure for a limited number of transmitting antennas. A further exploration of trellis coding for IM transmissions is presented in \cite{basar2011new}, where, in contrast to the approach in \cite{mesleh2010trellis}, trellis coding is employed in both the signal and spatial domains. Overall, the coded schemes studied in the previous works outperform the uncoded SM systems in practical bit error rate (BER) levels, yielding notable performance improvements across various channel conditions. More recent studies have explored the applicability of non-binary low-density parity-check codes \cite{feng2018nonbinary} and polar codes \cite{dai2021polar} to IM systems, as well as VLSI implementation for the iterative detection and decoding of coded IM signals\cite{liu2022maximum}.

Typically, IM transmission in the spatial domain requires one RF chain per active transmit antenna.  Activating a single antenna during transmission allows for the utilization of only one RF chain for the entire system \cite{mao2018novel}. On the other hand, an FA contains a single radiating liquid that moves along predetermined locations \cite{huang2021liquid}. Since only one radiating element exists, a single RF chain is sufficient and the FA-based system thus utilizes all its available resources. Consequently, index-based modulation can be applied to FAs by exploiting the single RF chain property. Besides, mutual coupling is a prevalent phenomenon in conventional antennas\cite{clerckx2007impact}. In the case of FAs, where a single element dynamically switches positions, the impact of mutual coupling is effectively eliminated. Recently, a transmission mechanism for FA-based MIMO communication systems based on index modulation was also studied, whereby a low-complexity detector was proposed by exploiting the sparsity of the transmitted FA-based signal vectors \cite{zhu2023index}.

Motivated by the above, this paper aims to exploit the use of IM transmission techniques in the context of FAs, while incorporating channel coding strategies to boost their performance. Since spatial correlation might be more detrimental to the spatial bits than the signal bits, we first present a set partition coding (SPC) scheme applied to the spatial domain to spatially separate the FA ports at each transmission time. While the analysis presented in \cite{mesleh2010trellis} showcases the benefits of utilizing IM transmissions with SPC, it does not delve into the scalability of the system to handle a large number of transmitting indices, a notable characteristic of FAs. Building on this, we present a turbo-coded modulation design for IM-FAs, which combines the extraordinary coding gain offered by turbo codes with IM transmissions. Thus, the usage of channel coding techniques for IM transmissions is to reduce the effects of spatial correlation imposed on the channel, especially for FAs that operate in confined spaces. We analyze the performance of the aforementioned scheme both at high signal-to-noise ratios (SNRs), by considering the overall weight enumerator function (WEF) of the turbo code, and at low SNRs, by exploring the convergence properties of iterative decoding through EXtrinsic Information Transfer (EXIT) charts. To the best of our knowledge, this is the first work that presents a complete framework for coded index-based modulation in the context of FA systems. Thus, to summarize, the contributions of this paper are threefold:

\begin{itemize}
\item The paradigm of index-based modulation within the framework of FA systems is introduced. In this approach, the fluidic element located within the FA is repositioned according to an index mapping table. We analyze the performance of the IM-FA system in terms of BER under correlated fading conditions. Our findings notably reveal that, particularly in confined operational space, the FA-based IM system outperforms the conventional IM antenna system, which is inherently associated with electromagnetic coupling.
\item We build on the previous model by studying a general SPC scheme for the space domain to spatially separate the FA ports. Specifically, the BER of the scheme is quantified by providing a tight upper bound directly related to the WEF of the convolutional code. Accordingly, this technique succeeds in reducing the correlation among FA ports, especially when the number of ports is large, and can thus provide large performance gains.
\item The spatial SPC scheme is extended to encompass the signal domain by employing a turbo-coded modulation design. The performance of turbo-coded IM-FAs is analyzed for low and high SNR regimes, wherein we show that employing error protection for both spatial and symbol bits provides significant performance gains over spatial SPC and uncoded systems. Indeed, our theoretical and simulation results reveal that, under fixed spectral efficiency constraints, the use of coded modulation designs further enhances the performance of IM-FAs.
\end{itemize}

The rest of the paper is organized as follows. Section II describes the considered system model, while Section III analyzes the BER of the IM-FA system. Section IV provides the spatial SPC scheme for IM-FAs. Section V introduces the turbo-coded modulation design for IM-FAs. Finally, Section VI presents the numerical results and Section VII concludes the paper.

\textit{Notation:} Lower and upper case boldface letters denote vectors and matrices, respectively; $[\cdot]^T$, $[\cdot]^H$ and ${\rm{tr}}(\cdot)$ denote the transpose, hermitian, and trace operators, respectively; $|\cdot|$ is the absolute value of a complex scalar; $\binom{a}{b}=\frac{a!}{b!(a-b)!}$ denotes the binomial coefficient; $\lfloor \cdot \rfloor$ is the floor function; $\mathds{1}_X$ is the indicator function of X with $\mathds{1}_X=1$, if $X$ is true, and  $\mathds{1}_X=0$, otherwise; $\log\left( \cdot \right)$ stands for the natural logarithm; $\mathcal{CN}(\mu,\sigma^2)$ denotes a complex Gaussian random variable with mean $\mu$ and variance $\sigma^2$; $f(\cdot)$ and $p(\cdot)$ represent the probability density function (PDF) and the probability mass function (PMF) of a random variable, respectively; $\mathbb{P}\{ \cdot \}$ and $\mathbb{E}\{ \cdot\}$ represent the probability and expectation operators, respectively; $H(\cdot,\cdot)$, and $I(\cdot;\cdot)$ denote the joint entropy and mutual information, respectively; $J_0(\cdot)$ is the zero-order Bessel function of the first kind; $Q\left(\cdot\right)$ is the Gaussian $Q$-function and $\mathcal{M}_{X}(\cdot)$ denotes the moment-generating function (MGF) of the random variable $X$. The main mathematical notations related to the system model are summarized in Table \ref{Table1}.

\section{System Model}
\begin{table*}[t]\centering
	\caption{Summary of Notations.}\label{Table1}
	\vspace{-1mm}
	\renewcommand{\arraystretch}{1.25}
	\centering
	\scalebox{0.88}{
		\begin{tabular}{| l | l || l | l |}\hline
			\textbf{Notation} & \textbf{Description} & \textbf{Notation} & \textbf{Description}\\\hline
			$N$ & Number of FA ports & $M$ & Modulation order \\\hline
			$\lambda$ & Wavelength of the carrier frequency&  $W$ & Length of the FA \\\hline
			$\mathbf{x}$ & Transmit vector & $\hat{{\bf{x}}}$ & Estimated vector \\\hline
			$d({\bf{x}},\hat{{\bf{x}}})$ & Hamming distance between vectors ${\bf{x}}$ and $\hat{{\bf{x}}}$ & $k$ & Spectral efficiency of the IM-FA \\\hline
			$k_{\rm{SPC}}$ & Spectral efficiency of the IM-FA with SPC & $k_{\rm{TCM}}$ & Spectral efficiency of the turbo-coded IM-FA\\\hline
			 $\mathcal{L}$ & Logarithm likelihood ratio & $\mathfrak{L}$ & Interleaver size\\\hline
	\end{tabular}}
	\vspace{-2mm}
\end{table*}

\begin{figure}
\includegraphics[width=0.485\textwidth]{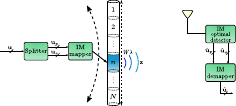}
\centering
\vspace{-2mm}
\caption{Transceiver structure for the proposed IM-FA system.}\label{fig:1}
\vspace{-3mm}
\end{figure}
We consider a point-to-point FA-based IM transmission system between a single FA transmitter and a conventional single-antenna receiver, as depicted in Fig. \ref{fig:1}. Specifically, a conductive fluid element is located within a uniform linear tube consisting of $N$ evenly distributed locations (also known as ports) along a linear dimension of $W\lambda$, where $\lambda$ is the transmission wavelength and $W$ is a scaling constant representing the size of the FA normalized by $\lambda$ \cite{wong2020fluid}. Moreover, the FA is equipped with a single RF chain and thus a single port is activated for transmission based on the output of the IM system. It is assumed that the conductive fluid element can switch locations instantly among the ports, e.g.,  with the assistance of a mechanical pump \cite{huang2021liquid}.

\subsection{Channel Model}
The signal is transmitted over a $1 \times N$ wireless channel $\bf{h}$. We consider a flat Rayleigh block fading communication channel where the channel coefficients remain constant during one timeslot, but change independently between different timeslots. Since the ports located within the FA structure are arbitrarily close to each other, the channels are considered to be correlated \cite{wong2020fluid}. Consequently, the correlated Rayleigh fading channel ${\bf{h}}$ is expressed as

\begin{equation}\label{eq:1}
{\bf{h}} = \widetilde{{\bf{h}}} \ {\boldsymbol{{\rm{R}}}}^{\frac{1}{2}},
\end{equation}
where the $N$ entries of $\widetilde{{\bf{h}}}$ are independent and identically distributed (i.i.d.) random variables, each following a complex circular Gaussian random variable with zero mean and variance $\sigma_h^2$. In addition, ${\boldsymbol{{\rm{R}}}}$ represents the $N \times N$ transmitter spatial correlation matrix. For the sake of presentation, the correlation matrix ${\boldsymbol{{\rm{R}}}}$ is written as

\begin{equation}\label{eq:2}
{\boldsymbol{{\rm{R}}}}=\left[\begin{array}{ccc}
R_{1,1} & \cdots & R_{1, N} \\[1mm]
\vdots & \ddots & \vdots \\[1mm]
R_{N, 1} & \cdots & R_{N, N}
\end{array}\right],
\end{equation}
where the $(i,j)$-th entry of ${\boldsymbol{{\rm{R}}}}$ is given by \cite{new2022fluid}

\begin{equation}\label{eq:3}
R_{i,j}= J_{0}\left(2 \pi\frac{(i-j)}{N-1}W\right),
\end{equation}
which characterizes the correlation between ports $i$ and $j$.

\subsection{Index-Based Modulation for FAs}
At the FA transmitter, the information bits are simultaneously sent over two parallel streams. The first one is mapped onto the spatial constellation diagram of size $N$, responsible for the port selection\footnote{In the context of IM-FA transmission, it is essential to highlight that the port selection is performed randomly and lacks adherence to specific criteria. This randomness stems from the inherent nature of the generated bit stream.}, while the second stream is mapped onto the signal constellation diagram of size $M$, responsible for the signal modulation. More specifically, the two parallel bit sequences ${\bf{u}}_p = [u_1, \ldots, u_{\tilde{n}}]$ of size $\tilde{n}=\log_2 N$, and ${\bf{u}}_s = [u_{\tilde{n}+1}, \ldots, u_{k}]$ of size $\tilde{m}=\log_2 M$ form together a bit sequence ${\bf{u}}=[u_1,\ldots, u_{k}]$ of size $k$. Notably, $N$ is restricted to a power of two since the FA port indices are mapped to binary blocks. Therefore, the spectral efficiency of the IM-FA system is $k = \tilde{n} + \tilde{m} = \log_2(NM)$ bits per second per Hertz (bps/Hz). Consequently, these $\log_2(NM)$ bits are mapped onto a constellation vector $\bf{x}$ of size $N$, whereby only one element in $\bf{x}$ is non-zero, i.e., at the position of the selected port. It follows that the activated port constitutes a mapping strategy, and the output vector is
\begin{align}\label{eq:4}
\mathbf{x} \triangleq\left[\begin{array}{lllllll}
0 & \cdots & 0 & \undermat{\!\!\!n{\text {-th}} \text { position}} \ \, \, \ {s_m} \ \ \ \, & 0 & \cdots & 0
\end{array}\right]^T,
\end{align}
\vspace{0.5mm}

\noindent where $n=1,2,\ldots,N$, represents the index of the activated port, and $s_m$ is the $m$-th information-bearing symbol from the $M$-ary constellation at the $n$-th position. In this work, we consider an $M$-ary quadrature amplitude modulation (QAM) constellation design in which $s_m \in \mathcal{S}$, where $\mathcal{S}$ denotes the QAM alphabet set. In contrast to conventional IM systems in which a single antenna is selected during transmission, FA-based IM systems allocate the radiating fluid element to a port determined by the IM mapper following the incoming bit stream during each transmission\footnote{A setup with multiple active ports could be implemented, yet would necessitate the use of multiple FAs, thereby introducing additional complexity.}. We note that while FAs share some conceptual similarities with conventional IM systems, FAs can support a much larger number of ports, and are not limited to a fixed physical structure \cite{huang2021liquid}.

The received signal experiences additive white Gaussian noise (AWGN) with component $w$ following a circularly symmetric Gaussian distribution with zero mean and variance $\sigma_{w}^2$, i.e., $w \sim \mathcal{CN}\left(0, \sigma_{w}^2\right)$. We assume a complex baseband signal representation and symbol-by-symbol detection in which the sampled signal at the receiver when the signal is transmitted from the $\! n$-th FA port is
\begin{equation}\label{eq:5}
y = h_n s_m + w,
\end{equation}
where $h_n$, $n=1,2,\ldots, N$, is the channel between the $n$-th FA port and the receiver, i.e., the $n$-th element of \eqref{eq:1}. It is assumed that the receiver has perfect knowledge of the channel \cite{wong2020fluid}. In this work, we consider a joint detection of the modulated symbol and the active FA port index, which is known to be the optimal detection \cite{jeganathan2008spatial}. Thus, the receiver operates based on the maximum likelihood (ML) criterion. Therefore, the detection reduces to
\begin{equation}\label{eq:6}
[\hat{n}, \hat{m}] = \underset{n,m}{\arg \min} \ \left| y - h_n s_m \right|^2,
\end{equation}
where $\hat{n}$ and $\hat{m}$ represent the indices of the estimated FA port and the symbol, respectively The ML is an optimal detector with search complexity that grows as $NM$ \cite{jeganathan2008spatial}. Other sub-optimal schemes with lower complexity could also be considered. Finally, the estimated indices for the port and symbol are de-mapped to obtain an estimate $\hat{{\bf{u}}}$ of the original information bit sequence $\bf{u}$.

\section{BER Analysis of IM-FAs}
In this section, we consider the analysis of IM-FAs in terms of BER. Due to the specific signal structure of IM systems, the transmit vector ${\bf{x}}$ is correctly recovered if both the port and the transmitted symbol are correctly detected. The operation of finding the active port is equivalent to solving an $N$-hypothesis testing problem at the receiver, i.e., the analysis involves the computation of multidimensional integrals. Hence, it is very common in the literature to compute the average BER by exploiting union-bound methods with respect to the pairwise error probability (PEP)\cite{mao2018novel}. We first provide in the following Lemma the corresponding PEP which facilitates the analysis in subsequent sections.

\textbf{Lemma 1.} \textit{The unconditional PEP that the transmitted vector $\bf{x}$ is received as another vector $\hat{\bf{x}}$ for IM-FAs is computed in a closed-form expression as}
\begin{align}\label{eq:8}
\mathbb{P} \left\{ \bf{x} \rightarrow \hat{\bf{x}} \right\} = \frac{1}{2}\left( 1 - \sqrt{\frac{1}{1 + \frac{4\sigma_w^2}{\mu}}} \right),
\end{align}
where $\mu$ is the eigenvalue of the matrix $\boldsymbol{\delta} \boldsymbol{\delta}^H {\boldsymbol{{\rm{R}}}}$, and $\boldsymbol{\delta} \triangleq  ({\bf{x}} - \hat{\bf{x}})$.
\begin{proof}
See Appendix A.
\end{proof}

It is important to note that the method employed to calculate the BER possesses a high degree of generality in the sense that any correlation model can be adopted, since the expression in \eqref{eq:8} depends solely on the eigenvalue of $\boldsymbol{\delta} \boldsymbol{\delta}^H {\boldsymbol{{\rm{R}}}}$. Hence, the entries of ${\boldsymbol{{\rm{R}}}}$ can follow any arbitrary model. We aim to provide a bound on the average BER. By using the result from Lemma 1, the average BER of IM-FAs is obtained as

\begin{align}\label{eq:9}
{\bar{P}}_{\rm{e}} & \leq \frac{1}{k 2^{k}} \sum_{{\bf{x}}} \sum_{\hat{{\bf{x}}}} d({\bf{x}},\hat{{\bf{x}}}) \mathbb{P} \left\{{\bf{x}} \rightarrow \hat{{\bf{x}}}\right\} \nonumber \\[1mm]
&= \frac{1}{2k 2^{k}} \sum_{{\bf{x}}} \sum_{\hat{{\bf{x}}}} d({\bf{x}},\hat{{\bf{x}}}) \left( 1 - \sqrt{\frac{1}{1 + \frac{4\sigma_w^2}{\mu}}} \right),
\end{align}
where $d({\bf{x}},\hat{{\bf{x}}})$ is the Hamming distance, i.e., the number of bits in error between the vectors ${\bf{x}}$ and $\hat{{\bf{x}}}$ and $k=\log_2(NM)$ is the number of transmitted bits.

\section{Spatial SPC for IM-FAs}
In this section, we build on the previous model in an effort to further enhance the performance. To this end, the goal is to minimize the spatial correlation between FA ports by applying the concept of SPC, first introduced in \cite{ungerboeck1982channel}, on the spatial domain. In particular, the set of transmitting port indices is partitioned into subsets such that the spacing between ports within a specific subset is maximized. For example, consider a 16-port FA where each port index is represented by 4 information bits. The role of the encoder is to select an appropriate subset that maximizes the Euclidean distance between consecutive bit sequences, effectively maximizing the spacing between ports. In other words, port indices that differ only at the most significant bit, such as $0000$ (port 1) and $1000$ (port 9), are grouped together. This encoding procedure places some restriction on the spatial bits that can be in any given sequence, thereby reducing spatial correlation. The working principle of IM-FAs with spatial SPC is provided next.

\begin{figure*}
\includegraphics[width=0.80\textwidth]{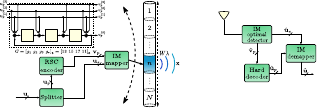}
\centering
\caption{Spatial SPC design for the IM-FA.}
\label{fig:2}
\end{figure*}
\subsection{Spatial SPC Design}
The incoming bit sequence ${\bf{u}}$ is split into two streams ${\bf{u}}_p$ and ${\bf{u}}_s$ similar to the uncoded case, as shown in Fig. \ref{fig:2}. To ensure that the FA ports are selected with maximum spatial separation, the port index bits ${\bf{u}}_p$ need to be processed by a recursive systematic convolutional (RSC) encoder of rate $(\tilde{n}-1)/\tilde{n}$, where $\tilde{n}=\log_2(N)$. Then, the output of the RSC encoder ${\bf{v}}_p$ along with the signal constellation bits go through the IM mapper to be transmitted over the channel. Consequently, the spectral efficiency of the IM-FA SPC scheme is 

\begin{equation}\label{eq:10}
k_{\rm{SPC}}= \log_2(MN)-1=k -1 \text{ bps/Hz},
\end{equation}
where $k=\log_2(NM)$ is the spectral efficiency of the uncoded system. Therefore, the coded FA system requires a higher modulation order to achieve the same spectral efficiency as the uncoded system (i.e., of Section III), due to the RSC code applied on the port index domain. In the case of coded modulation design, the output of the encoder is a succession of coded IM symbols, which for a sequence of length $\mathfrak{L}$ is given by

\begin{equation}\label{eq:11}
{\bf{x}} = \left[  x^{(1)}, x^{(2)}, \ldots, x^{(\mathfrak{L})} \right],
\end{equation}
where $x^{(l)}=\left(h_n^{(l)},s_m^{(l)}\right)$, $l=1,\ldots,\mathfrak{L}$, is the transmitted IM symbol at the $l$-th transmission interval, i.e., $s_m^{(l)}$ is the $m$-th symbol transmitted from the $n$-th FA port. At the receiver, the optimal ML sequence estimator (MLSE) is applied to jointly estimate the transmitting port index and the modulated symbol sequences. Then, the estimated port index bits $\bf{\hat{v}}_p$ first go through a hard decision Viterbi decoder, as depicted in Fig. \ref{fig:2}. Finally, the output bits $\bf{\hat{u}}_p$ from the Viterbi decoder together with the estimated symbol $\bf{\hat{u}}_s$ are used to recover the original information bits.

\subsection{BER Analysis of the IM-FA system with SPC}
In general, the BER analysis of IM-FAs with SPC can be carried out using the same method as that of convolutional codes, whereby the encoder WEF is obtained under the assumption of a memoryless binary symmetric channel (BSC). The BSC assumption holds whenever the bits are interleaved under perfect or infinite interleaving depth. Hence, the average BER of IM-FAs with SPC is upper bounded by \cite[Sec. 8.2.2]{massoud2007digital}
\begin{equation}\label{eq:12}
{\bar{P}}_{\rm{e}}^{\rm{SPC}} \leq \left. T(Z)\right\rvert_{Z=\zeta(d)} = \sum_{d=d_{\rm{free}}}^{\infty} a_d \zeta(d),
\vspace{-2mm}
\end{equation}
where $T(Z)$ is the WEF of the convolutional code which is represented by a polynomial that enumerates all codeword sequences of the form $Z^d$. Furthermore, the free distance of the code is denoted by $d_{\rm{free}}$, which represents the smallest Hamming distance between any two valid codewords. The term on the right-hand side is the infinite series representation of the WEF, where $a_d$ denotes the weight coefficient of a certain path of length $d$. Accordingly, for hard decision decoding, the term $\zeta(d)$ can be expressed as
\begin{align}
\zeta(d) =& \sum_{\kappa= \lfloor d/2 \rfloor+1}^d \binom{d}{\kappa} \mathcal{D}^\kappa\left(1-\mathcal{D}\right)^{d-\kappa} \nonumber \\
& \qquad \qquad \quad + \frac{1}{2}\binom{d}{d/2} \mathcal{D}^{d/2}\left(1-\mathcal{D}\right)^{d/2} \mathds{1}_{d \textnormal{ even}}.
\vspace{-2mm}
\end{align}
The term $\mathcal{D}$ is similar to the IM-FA system average BER derived in (\ref{eq:9}). However, the PEP term now deals with a sequence of IM symbols due to the interleaver. In other words, this term depends on the specific technique used to convey the information from the transmitter to the receiver. In the following proposition, we provide an expression for the PEP of the IM-FA system with SPC.

\textbf{Proposition 2.} \textit{The PEP that a sequence $\bf{x}$ of length $\mathfrak{L}$ is received as another sequence $\hat{\bf{x}}$ for an IM-FA system with SPC is given by
\begin{equation}\label{eq:14}
\mathbb{P} \left\{ \bf{x} \rightarrow \hat{\bf{x}} \right\} = \frac{1}{\pi} \int_{0}^{\pi/2} \prod_{l=1}^{\mathfrak{L}}\left( 1+ \frac{\mu^{(l)}}{4\sigma_w^2 \sin^2{\theta}} \right)^{-1} {\rm{d}}\theta,
\end{equation}
where $\mu^{(l)}$ is the eigenvalue of the matrix $\delta^{(l)} {\delta^{(l)}}^H {\boldsymbol{{\rm{R}}}^{(l)}}$, and $\delta^{(l)} \triangleq  {x^{(l)}} - \hat{x}^{(l)}$.}
\begin{proof}
See Appendix B.
\end{proof}

\begin{figure*}[t]
\includegraphics[width=0.79\textwidth]{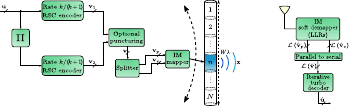}
\centering
\caption{Turbo-coded design for the considered IM-FA system.}\label{fig:3}
\vspace{-2mm}
\end{figure*}

\begin{figure*}[b]
\hrulefill
\begin{subequations}
\begin{align}
T(Z) &= \frac{5 Z^4 + 16 Z^5 - 12 Z^6 - 32 Z^7 + 4 Z^8 + 5 Z^9 + 29 Z^{11} - 
 25 Z^{13} + Z^{15} + 2 Z^{17}}{1 - 4 Z - 4 Z^2 + 4 Z^4 + 17 Z^5 + Z^7 - 9 Z^9 + 5 Z^{11} + 2 Z^{13}} \label{eq:subeq1}\\[2mm]
&= 5 Z^4 + 36Z^5 + 152Z^6 + 720Z^7 + 3472Z^8 +... \label{eq:subeq2}\\[2mm]
&= \sum_{d=4}^{\infty}a_d Z^d\label{eq:subeq3}
\end{align}
\end{subequations}
\end{figure*}

By using the result of Proposition 2, $\mathcal{D}$ can be expressed as \cite[Sec. 12.1]{simon2005digital}
\begin{align}\label{eq:15}
\mathcal{D} & \leq \frac{1}{k_{\rm{SPC}}} \sum_{{\bf{x}}} \sum_{\hat{{\bf{x}}}} d({\bf{x}},\hat{{\bf{x}}}) \mathbb{P} \left\{{\bf{x}} \rightarrow \hat{{\bf{x}}}\right\},
\end{align}
where $k_{\rm{SPC}}$ is the number of information bits given by \eqref{eq:10}. In what follows, for the sake of presentation, we adopt a rate $3/4$ RSC encoder for an FA with $N=16$ ports, as depicted in Fig. \ref{fig:2}. The encoder design is based on tabulated parameters taken from \cite[Table IV]{i2004design}. Note that existing studies that consider SPC with IM transmissions are typically restricted to a predetermined code structure \cite{mesleh2010trellis}. Thus, a significant challenge in using SPC for IM-FAs is adapting to higher code rates due to the large number of FA ports, which complicates the task of determining the WEF of the code. Additionally, while our design is not restricted to this specific encoder, the analysis becomes intractable as the number of encoder states increases. Nevertheless, the WEF is obtained from its weight-labeled state diagram to express each memory state as a function of the other state, and, hence, obtain the state equations. Upon solving these equations, a closed-form expression for the WEF of the considered RSC code is formulated and provided by \eqref{eq:subeq1} at the bottom of the page. Given that the error rate expression necessitates the weighing coefficients $a_d$, the ratio of polynomials in $Z$ is reformulated as an infinite series representation in \eqref{eq:subeq2}. Ultimately, the coefficients $a_d$ are used to obtain a performance bound for the IM-FA system with SPC as given in \eqref{eq:12}.

\section{Turbo-Coded IM-FAs}
In the previous section, a coded design was solely applied to the spatial domain to minimize the effects of the strong spatial correlation among the FA ports. However, the bits responsible for the signal constellation remained unprotected. Motivated by this, and in an attempt to further boost the performance of index-based modulation for FAs, we consider in this section a comprehensive coding scheme that covers both spatial and signal domains by employing a turbo-coded modulation design. It is important to mention that applying turbo codes for index-based modulation provides greater flexibility in selecting a suitable code rate \cite{wang2023road}. The consideration of other channel codes is an interesting direction for future work.

\subsection{Turbo-Coded Design}
At each transmission time, the transmitter processes a bit sequence $\mathbf{u}=\left[u_1, \ldots, u_k\right]$, as illustrated in Fig. \ref{fig:3}. Each sequence $\mathbf{u}$ is fed into a parallel-concatenated convolutional encoder, which uses a length $\mathfrak{L}$ symbol-based interleaver denoted by $\Pi$. Both component encoders illustrated in Fig. \ref{fig:3} are identical RSC encoders of rate $k / (k + 1)$. The output sequence of the first RSC encoder is denoted by $\mathbf{v}_1 = [\mathbf{u}, p_1]$, where $p_1 \in \{0, 1\}$ is the $(k+1)$-th bit of $\mathbf{v}_1$, representing the parity bit generated from $\mathbf{u}$ by the encoder. On the other hand, the output of the second RSC encoder is denoted by $\mathbf{v}_2=\left[\mathbf{u}, p_2\right]$, where $p_2 \in\{0,1\}$ is the $(k+1)$-th bit of $\mathbf{v}_2$. Accordingly, the output of the channel encoder is denoted by $\mathbf{v}=[\mathbf{u}, p_1, p_2]$. Thus, the overall turbo code rate is calculated as \cite[Sec. 6.1]{berrou2010codes}
\begin{equation}\label{eq:17}
R_c = \frac{k/(k+1)}{2-k(k+1)}=\frac{k}{k+2},
\end{equation}
where $k=\log_2\left({NM}\right)$. Therefore, the spectral efficiency of the turbo-coded IM-FA system is 
\begin{equation}\label{eq:18}
k_{\rm{TCM}}= \log_2(MN)-2=k -2 \text{ bps/Hz}.
\end{equation}
Finally, the IM mapper determines the activated port index and the modulated symbol using $\mathbf{v}$. Specifically, after splitting $\bf{v}$, the sequences ${\bf{v}}_p=\left[v_1, \ldots, v_{\tilde{n}}\right]$ and ${\bf{v}}_s=\left[v_{\tilde{n}+1}, \ldots, v_{k}\right]$ are mapped to the port index and the data symbol, respectively.

At the receiver, the demodulator first computes the \textit{a posteriori} logarithm likelihood ratios (LLR) of the transmitted bits, which are subsequently used as inputs to a channel decoder. The LLRs from the channel decoder concerning the information and parity bits are then processed as extrinsic information in successive iterations by the iterative turbo decoder block, as depicted in Fig. \ref{fig:3}. This sequence of operations is reiterated for a predefined number of iterations. Without loss of generality, we assume that the bits encoded into the spatial domain and the bits encoded into the signal domain are independent and generated with equal probability due to the interleaver \cite{schlegel2015trellis}. By taking into account the specific signal structure of the IM bits, the computation of the LLRs for the $n{\text {-th}}$ port index bit is given by
\vspace{2mm}
\begin{equation}\label{eq:19}
\begin{aligned}
\mathcal{L}(v_{p}^{(n)}) &= \log \frac{\mathbb{P}\left\{v_{p}^{(n)} = 1 \mid y\right\}}{\mathbb{P}\left\{ v_{p}^{(n)} = 0 \mid y\right\}} \\[2mm]
&= \log \frac{\sum_{\mathcal{P}_{n,1}} \sum_{\mathcal{S}} \exp\left(- |y-h_ns_m|^2 / \sigma_w^2\right)}{\sum_{\mathcal{P}_{n,0}} \sum_{\mathcal{S}} \exp\left(- |y-h_ns_m|^2 / \sigma_w^2\right)},
\end{aligned}
\vspace{1mm}
\end{equation}
where $v_{p}^{(n)}$ is the $n$-th coded port index bit, $n=1,2,\ldots,\tilde{n}$, $\mathcal{P}$ denotes the port index set, and $\mathcal{P}_{n,0}$ and $\mathcal{P}_{n,1}$ represent the subsets from the port index set which have $0$ and $1$ at the $n$-th bit, respectively. Similarly, the LLRs of the $m{\text {-th}}$ signal constellation bit is
\vspace{2mm}
\begin{equation}\label{eq:20}
\begin{aligned}
\mathcal{L}(v_{s}^{(m)}) &= \log \frac{\mathbb{P}\left\{v_{s}^{(m)} = 1 \mid y\right\}}{\mathbb{P}\left\{ v_{s}^{(m)} = 0 \mid y\right\}} \\[2mm]
&= \log \frac{\sum_{\mathcal{S}_{m,1}} \sum_{\mathcal{P}} \exp\left(- |y-h_ns_m|^2 / \sigma_w^2\right)}{\sum_{\mathcal{S}_{m,0}} \sum_{\mathcal{P}} \exp\left(- |y-h_ns_m|^2 / \sigma_w^2\right)},
\end{aligned}
\end{equation}
where $v_{s}^{(m)}$ is the $m$-th coded signal bit, $m=1,2,\ldots,\tilde{m}$. Moreover, $\mathcal{S}$ denotes the symbol set, and $\mathcal{S}_{m,0}$ and $\mathcal{S}_{m,1}$ represent the subsets from the symbol set which have $0$ and $1$ at the $m$-th bit, respectively. Ultimately, the output LLRs of the estimated sequence $\hat{{\bf{v}}}$, denoted by $\mathcal{L}\left( \hat{{\bf{v}}} \right)$, are processed by the iterative turbo decoder to recover an estimate of the original transmitted sequence $\hat{{\bf{u}}}$.

\subsection{Mutual Information Analysis of IM-FAs}
Considering the working mechanism of IM transmissions, where two information symbols are transmitted and jointly decoded, joint consideration of spatial and signal symbols is needed when deriving the theoretical capacity. More specifically, the information bits undergo modulation in both the spatial symbol, $h_n$, and the signal symbol, $s_m$. The mutual information is consequently a measure of the information gained about both the spatial and signal constellation spaces $\mathcal{H}=\{h_1, h_2,\ldots,h_N\}$ and $\mathcal{S}=\{s_1,s_2,\ldots,s_M \}$, respectively, by knowing the received signal space $\mathcal{Y}=\{y_1,y_2,\ldots, y_{NM} \}$. Therefore, the channel capacity can be expressed as \cite{biglieri2005coding}
\begin{align}\label{eq:21}
C &= \underset{p(h_n),p(s_m)}{\operatorname{max}} \ \left\{I\left( h_n, s_m; y\right)\right\} \nonumber \\[2mm]
&= \underset{p(h_n),p(s_m)}{\operatorname{max}} \ \left\{H\left(h_n, s_m\right) - H\left( h_n, s_m \mid y\right)\right\},
\end{align}
where $p(h_n)$ and $p(s_m)$ are the PMFs of the spatial and signal symbol, respectively. We deduce from \eqref{eq:21} that a part of the transmitted information is lost and captured by the term $H\left( h_n, s_m \mid y\right)$, called the channel equivocation. To avoid maximization over the PMFs of the spatial and signal domains, we assume that the IM symbols $r_{n,m} \triangleq h_n s_m$ are transmitted with equal probabilities \cite[Sec. 3.5.2]{biglieri2005coding}. Therefore, we rewrite \eqref{eq:21} as
\begin{align}\label{eq:22}
C = H\left(h_n, s_m\right) - H\left( h_n, s_m \mid y\right).
\end{align}
In the following proposition, we compute the capacity of the considered IM-FA system.

\textbf{Proposition 3.} \textit{The capacity of an $N$-port IM-FA system is given by
\begin{align}\label{eq:25}
C =& \log_2 \left(NM\right) -\frac{1}{NM} \sum_{n=1}^{N} \sum_{m=1}^{M} \mathbb{E}_{\mathcal{Y}\mid \mathcal{H},\mathcal{S}}\left\{ \log_2 \sum_{u=1}^{N}  \sum_{v=1}^{M} \right. \nonumber \\[4mm]
& \left. \qquad \qquad \exp\left(\frac{-|h_ns_m + w - h_us_v|^2 + |w|^2}{2\sigma_w^2} \right) \right\},
\end{align}
where the expectation over $\mathcal{H}$ and $\mathcal{S}$ is numerically evaluated.}
\begin{proof}
See Appendix C.
\end{proof}

As we show in the sequel, to obtain more insight, we characterize the performance of the considered turbo-coded IM-FA system by splitting the analysis into two parts; namely a high SNR and a low SNR region. Indeed, the asymptotic performance is formalized by providing a BER bound using the WEF of the turbo code, whereas the behavior of turbo codes at low SNR is better understood by statistical analysis, i.e., the EXIT analysis \cite{ng2008near}.

\subsection{BER for High SNRs}
We now provide a tight bound on the average BER of turbo-coded IM-FAs. The foundations for this bound are similar to the WEF bound applied to convolutional codes, as was applied in the previous section for IM-FAs with SPC. However, the WEF bounds for turbo codes differ from the usual WEF bounds for convolutional codes. Notably, turbo code bounds require a term-by-term joint enumerator for all possible combinations of input weights and output weights of error events. To formalize this, we introduce the input-output WEF (IOWEF) of a convolutional block code as a monomial that contains all potential codeword sequences generated by the encoder, which is expressed as
\begin{equation}\label{eq:26}
\mathcal{B}^{\mathcal{C}}(W,U,Z) = \sum_{w} \sum_{u} \sum_{z} \mathcal{B}_{w,u,z}^{\mathcal{C}} W^w U^u Z^z,
\end{equation}
where, $\mathcal{B}_{w,u,z}^{\mathcal{C}}$ denotes the number of paths with an input Hamming weight $w$, systematic output Hamming weight $u$, and parity check output Hamming weight $z$. The IOWEF can be obtained from its weight-labeled state diagram. However, since turbo codes are essentially convolutional block codes, the input information sequence has a finite length, thus we adopt the augmented state diagram method, which illustrates error events that include more than one deviation from the all-zero state in the trellis. The solution to the state equations provides us with the IOWEF, given by \cite{chatzigeorgiou2009augmented}
\begin{equation}\label{eq:27}
g\left(W,U,Z,L\right)=\sum_{l=1}^{\infty}g_l\left( W,U,Z \right)L^l,
\end{equation}
where $g_l\left( W,U,Z \right)$ enumerates all codeword sequences having specific path length $l$, and is defined as
\begin{equation}\label{eq:28}
g_l\left( W,U,Z \right)\triangleq \sum_w \sum_u \sum_z \mathcal{B}_{w,u,z,l}^{\mathcal{C}} W^w U^u Z^z.
\end{equation}
Therefore, $\mathcal{B}^{\mathcal{C}}(W,U,Z)$ is equivalent to $g_l\left( W,U,Z \right)$ for an input information sequence of length $l=\mathfrak{L}$. In other words, the IOWEF of a convolutional block code is written as
\begin{equation}\label{eq:29}
\mathcal{B}^{\mathcal{C}}(W,U,Z)\triangleq g_{\mathfrak{L}}(W,U,Z).
\end{equation}
In the following proposition, we provide a tight bound for the BER of turbo-coded IM-FAs.

\textbf{Proposition 4.} \textit{The average BER for turbo-coded IM-FAs is expressed as
\begin{align}\label{BER_TCM}
{\bar{P}}_e^{\rm{TCM}} &\leq \frac{1}{k_{\rm{TCM}}\mathfrak{L}}\sum_{w} \sum_{u} \sum_{z} w \mathcal{B}_{w,u,z}^{\mathcal{T}} \mathbb{P} \left\{{\bm{\mathrm{x}}} \rightarrow \hat{{\bm{\mathrm{x}}}}\right\} \nonumber \\
&=\frac{1}{k_{\rm{TCM}}\mathfrak{L}}\sum_{w} \mathcal{B}_{w}^{\mathcal{T}}(U,Z)\mathbb{P} \left\{{\bm{\mathrm{x}}} \rightarrow \hat{{\bm{\mathrm{x}}}}\right\},
\end{align}
where $\mathbb{P} \left\{{\bm{\mathrm{x}}} \rightarrow \hat{{\bm{\mathrm{x}}}}\right\}$ and $k_{\rm{TCM}}$  are given by \eqref{eq:14} and \eqref{eq:18}, respectively. Moreover, $\mathfrak{L}$, and $\mathcal{B}_{w}^{\mathcal{T}}(U,Z)$ respectively represent the interleaver size and the WEF of the turbo code.}
\begin{proof}
See Appendix D.
\end{proof}

\subsection{EXIT for Low SNRs}
To assess the performance of the system at low SNR regimes, the approach consists of predicting the behavior of the iterative decoder by solely looking at the input/output relationship of an individual constituent decoder. Owing to the encoding structure of the turbo-coded IM scheme, the systematic and parity bits of an encoded symbol are transmitted together over the communication channel. This means that the extrinsic and the systematic information associated with each \textit{a posteriori} symbol probability at the output of a constituent decoder cannot be separated. Therefore, we assume that the extrinsic and systematic information are independent of each other, so that the extrinsic information may be extracted from the \textit{a posteriori} symbol probability \cite{ng2008near}. This simplifying assumption enables us to generate the corresponding EXIT chart. More specifically, the average extrinsic information $I_{E}(u)$ at the output of the decoder can be computed as
\begin{equation}\label{eq:34}
I_{E}(u) = k_{\rm{TCM}} - \frac{1}{\mathfrak{L}} \sum_{l=1}^{\mathfrak{L}} \mathbb{E}\left\{ \sum_{i=1}^{2^{k_{\rm{TCM}}}} \mathcal{L}_{e}^{l,i} \log_2 \left( \mathcal{L}_{e}^{l,i}\right)\right\},
\end{equation}
where $\mathcal{L}_{e}^{l,i}$ represents the $i$-th extrinsic information symbol at the $l$-th time interval measured at the output of the decoder. Consequently, the EXIT is generated by numerical evaluation of the expectation in \eqref{eq:34}.

\section{Numerical Results}
In this section, we provide analytical and simulation results to quantify the performance of IM-FAs. The uncoded, set partition-coded, and turbo-coded fluid-based IM systems are denoted by IM-FA, IM-FA-SPC, and TC-IM-FA, respectively. Moreover, the conventional antenna-based IM system is denoted by IM-CA. We assume a fixed operating frequency of $5$ GHz ($\lambda=6$ cm), unless stated otherwise. Throughout this section, we maintain a fixed spectral efficiency whenever we compare the aforementioned schemes.

\begin{figure}
\includegraphics[width=0.485\textwidth]{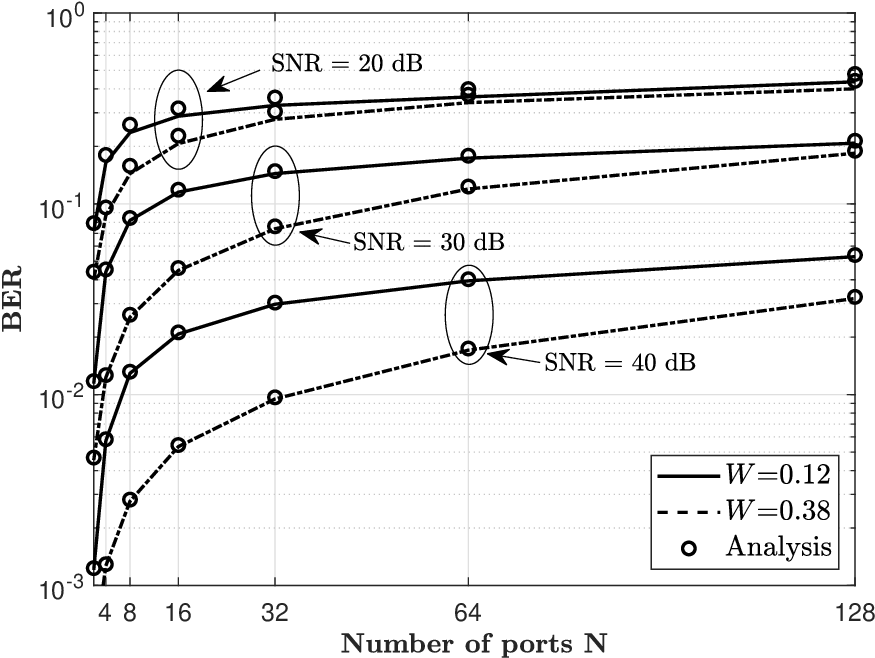}
\centering
\caption{Impact of the spatial correlation and the number of FA ports on the BER.}\label{fig:BER_vs_N}
\end{figure}

The performance in terms of uncoded BER of the IM-FA system for an increasing number of FA ports is shown in Fig. \ref{fig:BER_vs_N}. Specifically, we plot the results for two different lengths of the FA, i.e., $W=\{0.12, 0.38\}$, and set $M=4$. Firstly, we notice that for all SNR values, the BER performance gets worse as the number of FA ports increases. This is an expected result since the correlation between ports increases as well as the likelihood of erroneous port detection increases with an increasing number of FA ports. Secondly, we observe that for a lower number of FA ports, i.e., $N<64$, the BER increases significantly. Beyond this value, the performance does not degrade as much, meaning that it is sufficient to have a moderate number of FA ports since the BER is not largely affected. Lastly, we remark that the analytical results match with the simulations at high SNR regimes, which validates the derived bound.

\begin{figure}
\includegraphics[width=0.485\textwidth]{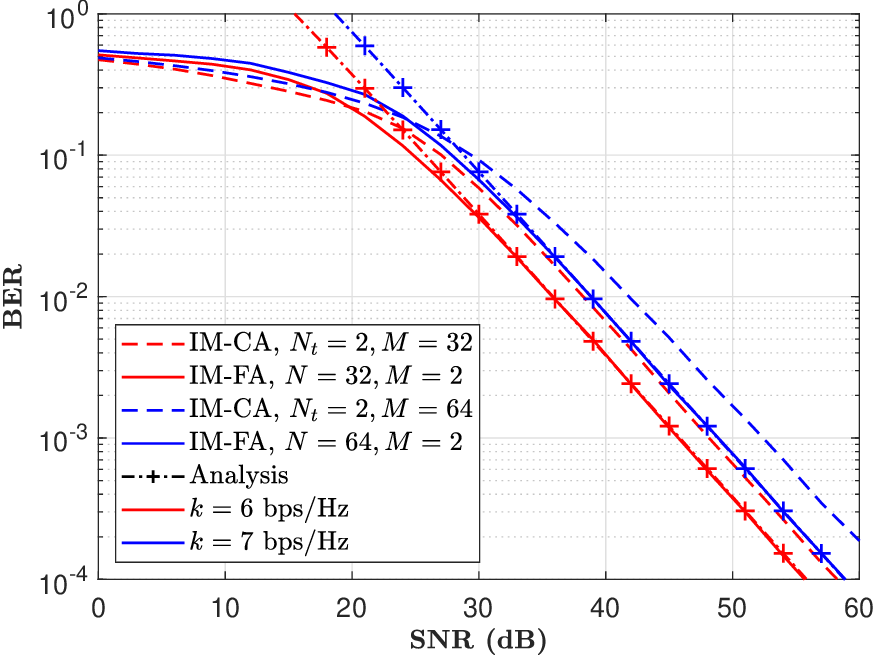}
\centering
\caption{Performance comparison between the IM-CA and IM-FA systems for different $k$.}\label{fig:IM_vs_IMFA}
\end{figure}

In Fig. \ref{fig:IM_vs_IMFA}, we provide a comparison between the IM-FA system and the IM-CA system with $N_t$ transmit antennas, for an uncoded scenario. To have a fair comparison, we choose a predefined spectral efficiency $k$ as well as the physical length of the FA. For instance, for $W=1$, the length of the FA is fixed at $6$ cm. Under this limitation, the IM-CA system can have at most $N_t=2$ transmit antennas if we want to keep them separated by at least half a wavelength. In addition, it should be noted that conventional IM transmissions are susceptible to antenna mutual coupling, resulting in performance degradation when the correlation is low. This inherent effect imposes an additional constraint when compared to FA-based IM transmission. Having this in mind, for $k=7$ bps/Hz, the IM-CA system requires a modulation order of $M=64$. On the other hand, the fluid-based IM system can accommodate at most $64$ ports with a minimum modulation order of $2$. We notice that the IM-FA system outperforms the $N_t$-antenna IM-CA system after a certain SNR value (around $24$ dB). More specifically, at a BER of $10^{-3}$, we observe a gain of approximately $4$ dB. This result is also validated for the case where $k=6$ bps/Hz. Finally, we observe that the analytical results match closely with the simulation results, which validates our theoretical framework.

\begin{figure}
\centering
\begin{subfigure}{.2445\textwidth}
  \centering
  \includegraphics[width=\linewidth]{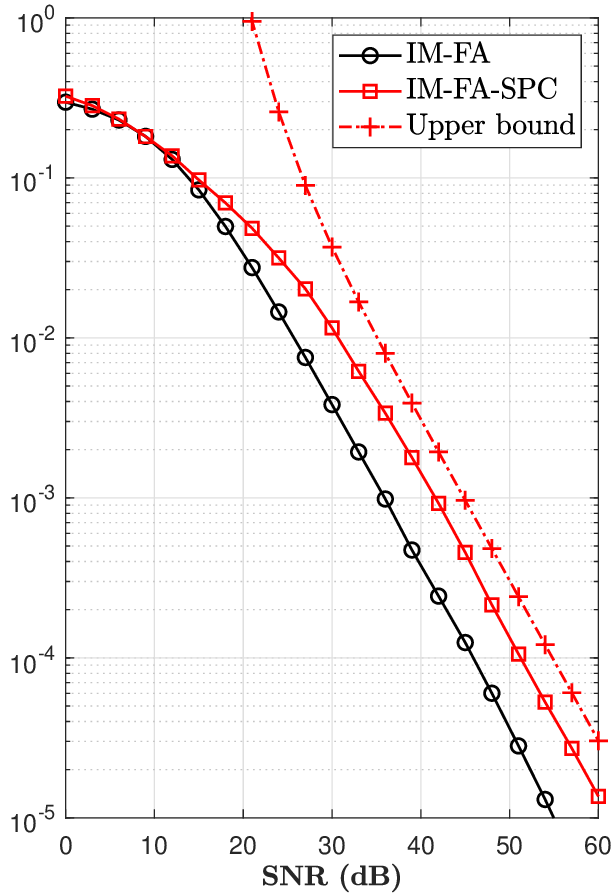}
  \caption{Low correlation ($W=5$)}
  \label{fig:sub1TCM}
\end{subfigure}%
\begin{subfigure}{.2445\textwidth}
  \centering
  \includegraphics[width=\linewidth]{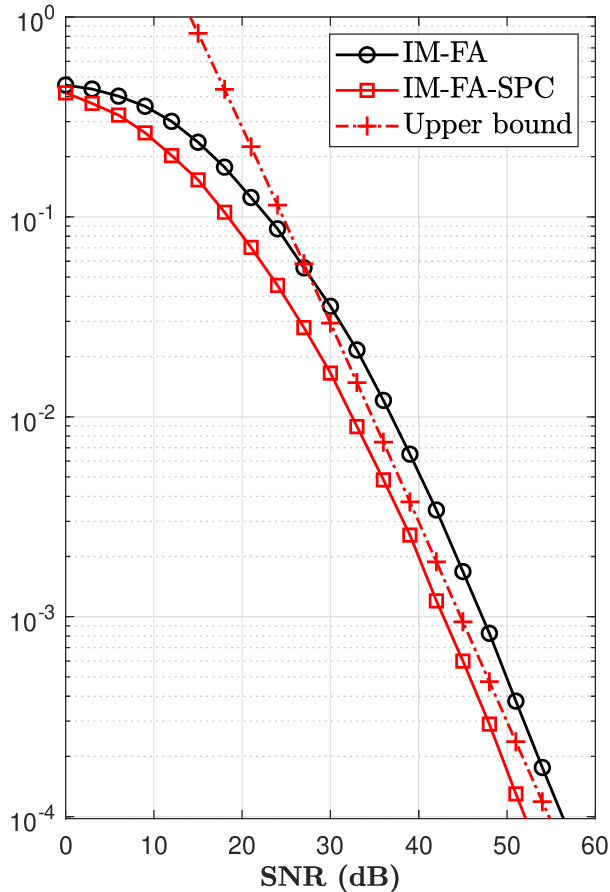}
  \caption{High correlation ($W=0.4$)}
  \label{fig:sub2TCM}
\end{subfigure}
\caption{Performance of IM-FAs with SPC for $N\!=\!16$, $k=5$ bps/Hz.}
\vspace{-2mm}
\label{fig:1D_TCM}
\end{figure}

We present the performance of IM-FAs with SPC in Fig. \ref{fig:1D_TCM} for the specific case where the FA is equipped with $N=16$ ports. To quantify the performance, we consider two scenarios for the correlation; the first with a large FA physical length, i.e., $W=5$, and the second with $W=0.4$ to generate a high correlation. Moreover, we consider the rate $3/4$ RSC code with generator polynomials $\left[g_1 \ g_2 \ g_3 \ g_r\right]_8 = \left[13 \ 15 \ 17 \ 11\right]_8$ as depicted in Fig. \ref{fig:2}. To achieve the same spectral efficiency of $k=5$ bps/Hz, the IM-FA system employs a modulation order of $M=2$, whereas the IM-FA-SPC scheme requires $M=4$. We first observe that for the low correlation scenario in Fig. \ref{fig:sub1TCM}, the use of SPC for the FA ports has no advantage and performs worse than IM-FAs with no SPC. The reason for this behavior is that the uncoded FA-based system uses a lower modulation order when compared to its coded counterpart, hence the expected coding gain is not visible whenever the spatial correlation is low. On the other hand, the second scenario in Fig. \ref{fig:sub2TCM} shows the advantage of IM-FA-SPC over the uncoded IM-FA system in the presence of high correlation, as it is apparent that we have a significant performance improvement in terms of coding gain of approximately $4.2$ dB at a BER of $10^{-4}$. The gain improvement is mainly attributed to the encoding structure and the set partitioning of the FA ports. In other words, as the spatial separation between ports increases, the effect of spatial correlation is reduced. On a final note, we account for the first 10 non-zero terms of the upper bound in \eqref{eq:12}. Accordingly, we observe that the derived theoretical bound for the IM-FA-SPC system follows the same trend as the simulation results, thereby validating our analysis.

\begin{figure}
\includegraphics[width=0.485\textwidth]{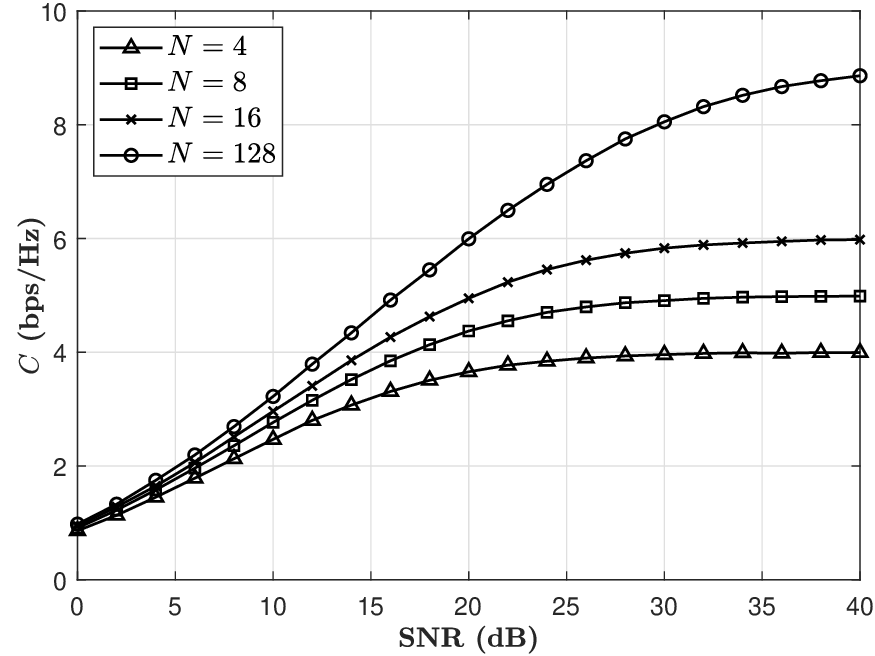}
\centering
\caption{Capacity of the considered IM-FA system for $M=4$ and $N \in \{4,8,16,128\}$ ports.}\label{fig:MI_analysis}
\end{figure}

The results of the discrete input capacity analysis for the IM-FA system, considering $M=4$ and varying values of $N \in \{4,8,16,128 \}$, are illustrated in Fig. \ref{fig:MI_analysis}. A noteworthy observation is that, for all considered values of $N$, the high SNR asymptotic performance is confined to $\log_2\left( NM \right)$. This limitation stems from the discrete nature of the FA ports and constellation order. To quantify the results, we consider that a transmission rate of $4$ bps/Hz is achieved by utilizing $M=4$ and $N=4$ FA ports. It is evident that error-free transmission can be attained at around $28$ dB. However, with a doubling of the number of FA ports, i.e., $N=8$, error-free transmission becomes feasible at $17.5$ dB. This observation highlights the significance of the number of FA ports in influencing the system's performance, particularly in achieving reliable communication at lower SNR levels.

\begin{figure}
\includegraphics[width=0.455\textwidth]{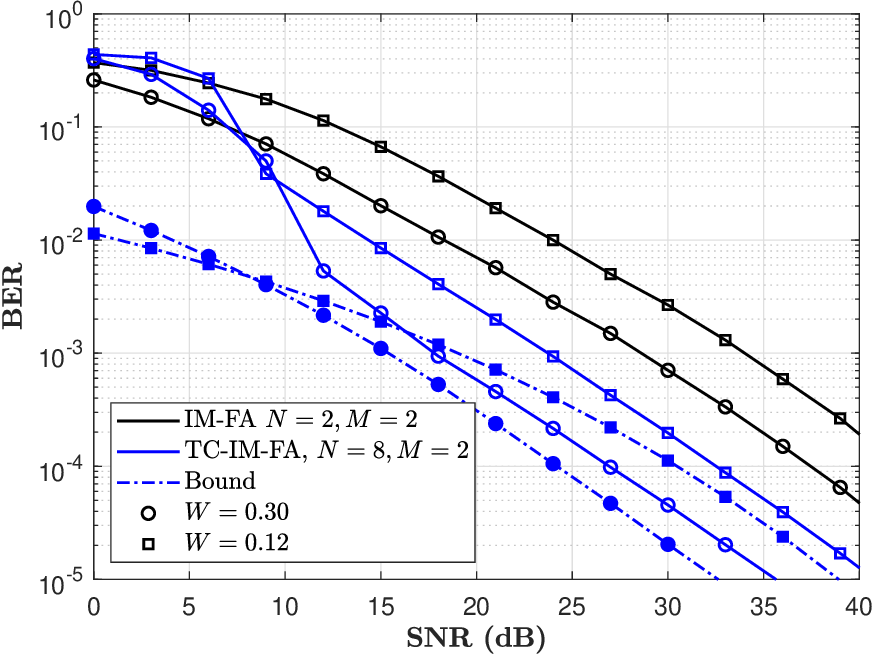}
\centering
\caption{BER performance of the TC-IM-FA scheme for $k=2$ bps/Hz.}\label{fig:TTCM_BER}
\end{figure}

We analyze in Fig. \ref{fig:TTCM_BER} the performance of the TC-IM-FA system for a fixed spectral efficiency of $2$ bps/Hz. The employed constituent RSC code in this configuration is an 8-state, rate $2/3$ code with $\left[g_1 \ g_2 \ g_r\right]_8 = \left[2 \ 4 \ 11\right]_8$ as generator polynomials\footnote{For the sake of simplicity, we use RSC codes with fewer memory states to keep the analysis tractable since having a larger memory size provides negligible gains regarding the onset of the turbo cliff region \cite[Sec. 10.8]{schlegel2015trellis}.}. The overall turbo code then operates at a rate of $2/4$, where $N=8$ and $M=2$ to satisfy \eqref{eq:18}. The interleaver size is set to 4096 symbols, and the number of turbo iterations is fixed at $15$. The performance of the considered TC-IM-FA scheme is compared with the uncoded IM-FA system, with $N=2$ FA ports and a modulation order of $M=2$ to ensure an equivalent spectral efficiency. The presented results consider two distinct FA sizes, denoted by $W=\{0.12,0.30\}$, representing different correlation scenarios. Upon examining the performance of the TC-IM-FA scheme in terms of the FA size, it is evident that the BER is lower for an FA size of $W=0.3$ compared to $W=0.12$. This difference is quantified at around $5$ dB at a BER of $10^{-5}$. In other words, this result showcases that a larger $W$ (indicative of a lower correlation among FA ports) exhibits a lower error floor, and as a result, achieves a better asymptotic performance. Another observation is that, at a BER of $10^{-4}$, the TC-IM-FA system remarkably outperforms the uncoded IM-FA system by approximately $9.8$ dB, particularly for $W=0.30$. This gain is attributed to the iterative message-passing procedure at the receiver. Lastly, we observe that the bounds of the TC-IM-FA scheme for both FA sizes exhibit an asymptotically good match concerning their corresponding BER curves, thereby validating the system performance at high SNR.

\begin{figure}
\includegraphics[width=0.455\textwidth]{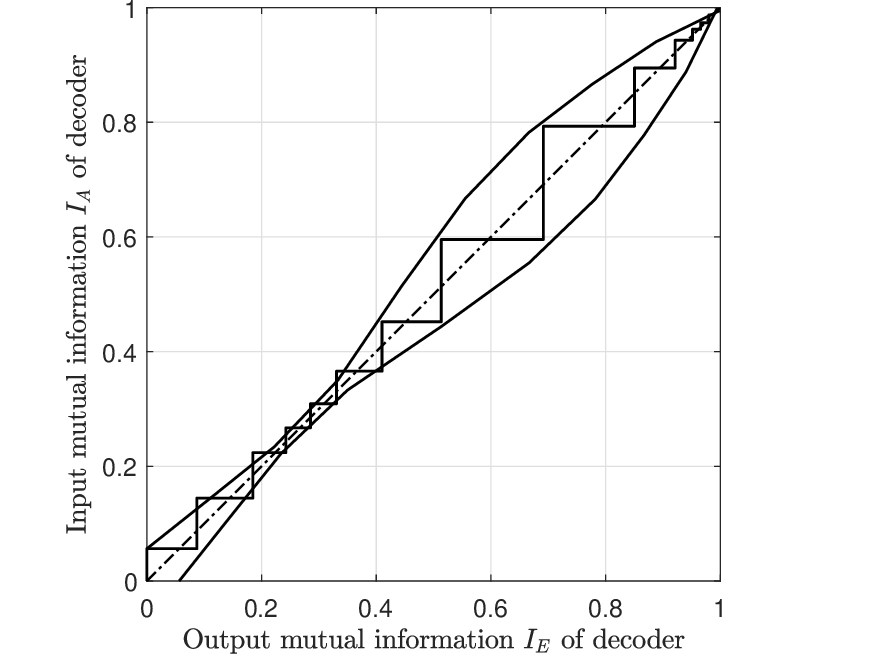}
\centering
\caption{EXIT chart analysis for the rate $2/3$, $8$-state, $\left[g_1 \ g_2 \ g_r\right]_8 = \left[2 \ 4 \ 11\right]_8$ RSC code.}\label{fig:TTCM_EXIT}
\end{figure}

\begin{figure*}
\centering
\begin{subfigure}{.5\textwidth}
  \centering
  \includegraphics[width=\linewidth]{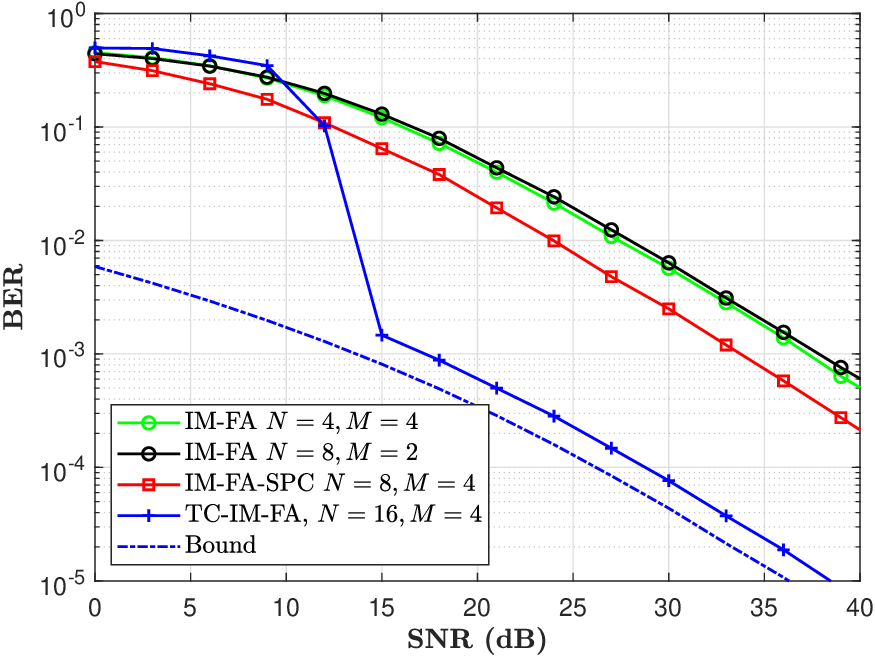}
\caption{$k=4$ bps/Hz.}\label{fig:All_schemes_BER_4bpcu}
\end{subfigure}%
\begin{subfigure}{.5\textwidth}
  \centering
  \includegraphics[width=\linewidth]{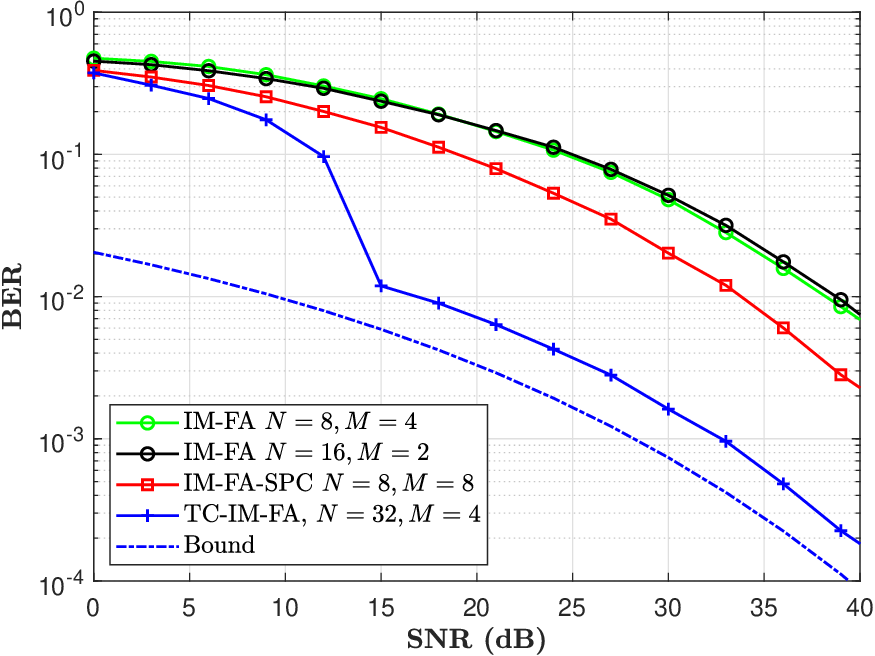}
  \caption{$k=5$ bps/Hz.}\label{fig:All_schemes_BER_5bpcu}
\end{subfigure}
\caption{Performance of IM-FA, IM-FA-SPC, and TC-IM-FA at different spectral efficiencies.}
\label{fig:All_schemes_BER}
\end{figure*}

To assess the performance of the TC-IM-FA system at low SNR, we illustrate in Fig. \ref{fig:TTCM_EXIT} the convergence characteristics of the corresponding turbo code employed in Fig. \ref{fig:TTCM_BER}. Specifically, we focus on an FA size of $W=0.12$, involving the 8-state, rate $2/3$ constituent RSC code with generator polynomials $\left[g_1 \ g_2 \ g_r\right]_8 = \left[2 \ 4 \ 11\right]_8$, using an interleaver length of $24,649$ symbols.  The input mutual information $I_A$ is plotted versus the output extrinsic mutual information $I_E$. Additionally, we depict the trajectory of the iterative decoding process at an SNR of $6.2$ dB, for a total of $15$ turbo iterations. The axis extends from 0 to 1 bit of mutual information, i.e., the measure has been normalized per coded bit \cite[Sec. 12.4]{schlegel2015trellis}. The opening observed in the trajectory corresponds to the start of the turbo cliff region in the BER curve illustrated in Fig. \ref{fig:TTCM_BER}. This means that after the sixth iteration through the decoder, fast convergence towards a BER floor is observed. It is essential to note that the steps in the trajectory might not perfectly align with the EXIT chart due to two main assumptions: first, the trajectory remains on the characteristics for additional passes through the decoder only in the case of an infinite-length interleaver, which in our case is not possible. Second, the EXIT charts were generated with the assumption that the extrinsic and the systematic information are independent, which has limited validity. Nevertheless, despite these considerations, the EXIT chart analysis offers a valuable tool for understanding the intricacies of iterative decoding processes.

In Fig. \ref{fig:All_schemes_BER}, we provide a performance comparison among the considered schemes for IM-FA systems for higher spectral efficiencies, i.e., for $k= 4$ and $5$ bps/Hz. For both scenarios, we employ an FA size of $W=1.0$ and assume an interleaver size of $1024$ symbols for the coded schemes. Moreover, we consider $10$ turbo iterations for the TC-IM-FA case. For the case of $k=4$ bps/Hz in Fig. \ref{fig:All_schemes_BER_4bpcu}, the IM-FA system uses $N=4$, $M=4$ and $N=8$, $M=2$. As for the IM-FA-SPC scheme, we use the $8$-state, rate $2/3$ RSC code with generator polynomials $\left[g_1 \ g_2 \ g_r\right]_8 = \left[2 \ 4 \ 11\right]_8$, i.e., $N=8$ with $M=4$. The TC-IM-FA scheme employs the $16$-state, rate $4/5$ constituent RSC code with polynomials $\left[g_1 \ g_2 \ g_3 \ g_4 \ g_r\right]_8 = \left[33 \ 27 \ 31 \ 23 \ 35\right]_8$. Hence, the overall turbo code is of rate $4/6$, i.e., $N=16$ ports ($4$ bits) and $M=4$ ($2$ bits). Furthermore, for the case of $k=5$ bps/Hz in Fig. \ref{fig:All_schemes_BER_5bpcu}, the same RSC code is used for the IM-FA-SPC scheme with a modulation order of $M=8$. As for the TC-IM-FA system, the constituent RSC code used is the $8$-state, rate $5/6$ code with polynomials $\left[g_1 \ g_2 \ g_3 \ g_4 \ g_5\ g_r\right]_8 = \left[5 \ 7 \ 11 \ 13 \ 15 \ 17\right]_8$. The overall turbo code is of rate $5/7$, i.e., $N=32$ ports ($5$ bits) and $M=4$ ($2$ bits). First, we notice for the IM-FA system that the interplay between the port index domain and the modulation order (black and green curves) minimally impacts performance. However, this finding introduces design flexibility depending on the application's specific requirements. Then, the evident observation from both scenarios is that the TC-IM-FA scheme exhibits a substantial BER reduction compared to the IM-FA-SPC and the uncoded IM-FA schemes. Notably, for $k=4$ bps/Hz, it provides nearly a 12 dB gain over the IM-FA-SPC at a BER of $10^{-4}$, primarily due to the turbo decoding process. Finally, when comparing the performance of the TC-IM-FA scheme in both scenarios, it is clear that the turbo cliff region is smaller when the spectral efficiency is increased. In other words, the BER reduction becomes less significant, and the performance of the turbo-coded case reaches an error floor at a higher BER. Nevertheless, this gain can be further enhanced by increasing the interleaver size and the memory of the constituent RSC code, given no constraints on latency and memory requirements.

\section{Conclusion}
In this work, we studied the concept of IM-FA systems, in which an FA transmitter's port indices are used to convey additional information bits. Initially, we derived a closed-form expression for the average uncoded BER of IM-FAs. Given that the FA operates in a small physical space, we employed channel coding techniques to further enhance the performance. Our results first demonstrated that by applying spatial SPC, we observed a coding gain improvement of 4 dB compared to the uncoded scenario. Then, by implementing a turbo-coded modulation design to encompass both spatial and signal domains, the performance of index-based modulation for FA systems is significantly improved, given fixed spectral efficiency constraints. Ultimately, this work has aimed to introduce coded IM transmission techniques for the emerging FA technology and has highlighted their influence in next-generation communication systems.
\appendices
\section{Proof of Lemma 1}
Based on the ML detection rule in \eqref{eq:6}, the conditional PEP is calculated as
\allowdisplaybreaks
\begin{align}
\mathbb{P} \left\{ \bf{x} \rightarrow \hat{\bf{x}}  \mid {\bf{h}} \right\} &= \mathbb{P} \left\{ | y - {\bf{h}} {\bf{x}} |^2 > | y - {\bf{h}} \hat{\bf{x}} |^2 \mid {\bf{h}}\right\} \nonumber \\[2mm]
&= \mathbb{P} \left\{ | w |^2 > | {\bf{h}}( {\bf{x}} - {\hat{\bf{x}}}) + w |^2 \mid {\bf{h}}\right\} \nonumber \\[2mm]
&= \mathbb{P} \left\{ \! | w | | {\bf{h}}( {\bf{x}} - {\hat{\bf{x}}}) | <  - \frac{1}{2} |{\bf{h}}( {\bf{x}} \! - \! {\hat{\bf{x}}}) |^2 \! \mathrel{\bigg|} \! {\bf{h}} \! \right\}. \label{eq:35}
\end{align}
Then, given $\bf{h}$, we obtain a Gaussian-distributed random variable with zero mean and variance $\sigma_w^2 | {\bf{h}}( {\bf{x}} - {\hat{\bf{x}}}) |^2$. Hence, $\mathbb{P} \left\{ \bf{x} \rightarrow \hat{\bf{x}}  \mid {\bf{h}} \right\}$ is reformulated as
\begin{align} \label{eq:36}
\mathbb{P} \left\{ \bf{x} \rightarrow \hat{\bf{x}}  \mid {\bf{h}} \right\} = Q\left( \sqrt{\frac{| {\bf{h}}( {\bf{x}} - {\hat{\bf{x}}}) |^2}{4\sigma_w^2}} \right).
\end{align}

The next step is to obtain the average PEP. We can rewrite the PEP  in \eqref{eq:36} as
\begin{equation}\label{eq:37}
\mathbb{P} \left\{ \bf{x} \rightarrow \hat{\bf{x}}  \mid {\bf{h}} \right\} = \frac{1}{\pi} \int_0^{\frac{\pi}{2}} \exp\left( -\frac{ | {\bf{h}}( {\bf{x}} - {\hat{\bf{x}}}) |^2}{8\sigma_w^2 \sin^2\theta} \right) {\rm{d}}\theta,
\end{equation}
where we used the following alternative (Craig's) representation of the Gaussian $Q$-function \cite[Sec. 4.1, Eq. (4.2)]{simon2005digital}
\begin{equation}\label{eq:38}
Q(z) = \frac{1}{\pi} \int_0^{\frac{\pi}{2}} \exp\left( -\frac{z^2}{2 \sin^2\theta} \right) {\rm{d}}\theta.
\end{equation}
Then, by taking the expectation of (\ref{eq:37}) over $\bf{h}$, we obtain
\begin{equation}\label{eq:39}
\mathbb{P} \left\{ \bf{x} \rightarrow \hat{\bf{x}} \right\} = \frac{1}{\pi} \int_0^{\frac{\pi}{2}} \mathcal{M}_{\Psi}\left( -\frac{\bar{\gamma}}{2\sin^2\theta} \right) {\rm{d}}\theta,
\end{equation}
where $\Psi=| {\bf{h}}( {\bf{x}} - {\hat{\bf{x}}}) |^2$, and $\bar{\gamma} \triangleq 1 / (4\sigma_w^2)$. In the special case where the channel coefficients are i.i.d., the MGF is decomposed into a product of marginal MGFs. However, in the presence of spatial correlation, this is not possible.
By expressing the MGF of the random variable $\Psi$ in terms of the correlation matrix, we obtain \cite{hedayat2005analysis}
\begin{equation}\label{eq:40}
\mathcal{M}_{\Psi}(s) = \prod_{n=1}^N \frac{1}{1-s \mu_n},
\end{equation}
where $\mu_n$ is the $n$-th eigenvalue of $\boldsymbol{\delta} \boldsymbol{\delta}^H {\boldsymbol{{\rm{R}}}}$. Note that the analysis, in this case, is simplified due to having a single antenna receiver. Significantly, we notice the matrix $\boldsymbol{\delta} \boldsymbol{\delta}^H {\boldsymbol{{\rm{R}}}}$ has always rank one due to the term $\boldsymbol{\delta} \boldsymbol{\delta}^H$, which is a consequence of employing IM transmission. Thus, we obtain a single non-zero eigenvalue, denoted by $\mu$, which can be calculated as
\begin{equation}\label{eq:41}
\mu = \sum_{n=1}^N \mu_n = {\rm{tr}}\left( \boldsymbol{\delta} \boldsymbol{\delta}^H {\boldsymbol{{\rm{R}}}} \right).
\end{equation}
Hence, the MGF simplifies to
\begin{equation}\label{eq:42}
\mathcal{M}_{\Psi}(s) = \frac{1}{1-s \mu}.
\end{equation}
Ultimately, by replacing the above expression in (\ref{eq:39}), we obtain a closed-form solution for the PEP of IM-FAs in (\ref{eq:8}).

\section{Proof of Proposition 2}
The received signal at time interval $l$ is written as
\begin{equation}\label{eq:43}
y^{(l)} = h_n^{(l)} s_m^{(l)} + w^{(l)},
\end{equation}
where $ h_n^{(l)} $ is the channel between the $n$-th FA port and the receiver at time instant $l$, $s_m^{(l)}$ is the information symbol, and $w^{(l)}$ is the AWGN component. Based on the observation over the entire sequence, the MLSE is applied to jointly estimate the activated port index and the modulated symbol, i.e.,
\begin{equation}\label{eq:44}
[\hat{{\bf{n}}}, \hat{\bf{m}}] = \underset{n,m}{\operatorname{arg \ min}} \ \sum_{l=1}^{\mathfrak{L}}\left| y^{(l)} - h_n^{(l)} s_m^{(l)} \right|^2,
\end{equation}
where $\hat{{\bf{n}}}=\left[ n^{(1)}, \ldots, n^{(\mathfrak{L})}\right]$ and $\hat{{\bf{m}}}=\left[ m^{(1)}, \ldots, m^{(\mathfrak{L})}\right]$.
The analysis follows closely to the adopted approach in Appendix A to compute the PEP. However, for coded scenarios, the decision is made based on an observation of the received signal for a sequence of length $\mathfrak{L}$. Thus, we can express the PEP in this case as
\begin{equation}\label{eq:45}
\mathbb{P} \left\{ \bf{x} \rightarrow \hat{\bf{x}} \right\} = \frac{1}{\pi} \int_0^{\frac{\pi}{2}} \prod_{l=1}^{\mathfrak{L}} \mathcal{M}_{\Psi^{(l)}}\left( -\frac{\bar{\gamma}}{2\sin^2\theta} \right) {\rm{d}}\theta,
\end{equation}
where $\Psi^{(l)}=\left| {{\bf{h}}^{(l)}}(x^{(l)} - {\hat{x}^{(l)}}) \right|^2$ and $\bar{\gamma} \triangleq 1 / (4\sigma_w^2)$. Asymptotically, an infinite-length sequence transforms the quasi-static fading channel into a fast fading channel. Hence, we consider the case where ${{\bf{h}}^{(l)}}$ are independent across $l$, but the entries of ${{\bf{h}}^{(l)}}$ are correlated. Consequently, the MGF is given by
\begin{equation}\label{eq:46}
\mathcal{M}_{\Psi^{(l)}}(s) = \frac{1}{1-s \mu^{(l)}}.
\end{equation}
Finally, the result follows by replacing the MGF term in \eqref{eq:45} to obtain the PEP expression as given in \eqref{eq:14}.

\section{Proof of Proposition 3}
The first term on the right-hand side of \eqref{eq:22} represents the maximum number of bits that can be transmitted, i.e.,
\begin{equation}\label{firstRHS}
H\left(h_n, s_m\right) = \log_2{(NM)}.
\end{equation}
Moreover, according to the channel model, the PDF of $y$, conditioned on $h_n$ and $s_m$ comes from a $\mathcal{CN}(h_n s_m,\sigma_{w}^2)$ distribution, and is given by
\begin{equation}\label{eq:23}
f\left( y \mid h_n, s_m \right) = \frac{1}{\sqrt{2\pi \sigma_w^2}}\exp\left(\frac{-| y-h_n s_m |^2}{2\sigma_w^2}\right).
\end{equation}
Consequently, by applying Bayes's rule, the second term on the right-hand side of \eqref{eq:22} is calculated as
\begin{align}\label{secondRHS}
& H\left( h_n, s_m \mid y\right) = \mathbb{E}_{\mathcal{H},\mathcal{S},\mathcal{Y}}\left\{ \log_2 \frac{1}{f(h_n,s_m \mid y)}  \right\} \nonumber \\[3mm]
& \ = \mathbb{E}_{\mathcal{H},\mathcal{S},\mathcal{Y}}\left\{ \log_2 \frac{\sum_{u=1}^{N} \sum_{v=1}^{M}f(y \mid h_u,s_v)}{f(y \mid h_n,s_m)}  \right\} \nonumber \\[3mm]
& \ = \frac{1}{NM} \sum_{n=1}^{N} \sum_{m=1}^{M} \nonumber \\[3mm]
& \qquad \mathbb{E}_{\mathcal{Y}\mid \mathcal{H},\mathcal{S}}\left\{ \log_2\left( \frac{\sum_{u=1}^{N} \sum_{v=1}^{M}\exp\left(\frac{-| y-h_u s_v |^2}{2\sigma_w^2}\right)}{\exp\left(\frac{-| y-h_n s_m |^2}{2\sigma_w^2}\right)} \right) \right\} \nonumber \\[3mm]
& \ = \frac{1}{NM} \sum_{n=1}^{N} \sum_{m=1}^{M} \mathbb{E}_{\mathcal{Y}\mid \mathcal{H},\mathcal{S}}\left\{ \log_2 \sum_{u=1}^{N} \sum_{v=1}^{M} \exp\left( \mathcal{A} \right) \right\},
\vspace{2mm}
\end{align}
where $\mathcal{A}$ is given by
\begin{equation}
\mathcal{A}=\frac{-|h_ns_m \! + \! w \!-\! h_us_v|^2 \!+\! |w|^2}{2\sigma_w^2}.
\end{equation}
The proof follows from combining \eqref{firstRHS} and \eqref{secondRHS} to obtain the final result in \eqref{eq:25}.

\section{Proof of Proposition 4}
Since turbo codes are a parallel concatenation of two convolutional block codes, the goal is to connect the codewords of the first constituent encoder to the codewords of the second. Having this in mind, we introduce conditional WEF (CWEF), which provides all codeword sequences given a specific input weight. Thus, the IOWEF of the convolutional block code in \eqref{eq:29} is rewritten as
\begin{equation}\label{eq:30}
\mathcal{B}^{\mathcal{C}}(W,U,Z)=\sum_w \mathcal{B}_w^{\mathcal{C}}(U,Z) W^w,
\end{equation}
where
\vspace{3mm}
\begin{equation}\label{eq:31}
\mathcal{B}_w^{\mathcal{C}}(U,Z)=\sum_u \sum_z \mathcal{B}_{w,u,z}^{\mathcal{C}} U^u Z^z,
\end{equation}
is the conditional WEF (CWEF) given an input weight $w$.
Furthermore, the CWEF of the turbo code $\mathcal{B}_{w}^{\mathcal{T}}(U,Z)$ can be obtained using the CWEFs of its constituent codes by assuming a uniform interleaver of size $\mathfrak{L}$. Specifically, we can express $\mathcal{B}_{w}^{\mathcal{T}}(U,Z)$ as \cite[Sec. 10.6]{schlegel2015trellis}
\vspace{3mm}
\begin{equation}\label{eq:32}
\mathcal{B}_{w}^{\mathcal{T}}(U,Z) = \frac{\mathcal{B}_{w}^{C_1}(U,Z) \ \mathcal{B}_{w}^{C_2}(U=1,Z)w! \left(\mathfrak{L}-w\right)!}{\mathfrak{L}!},
\vspace{1mm}
\end{equation}
where $\mathcal{B}_{w}^{C_1}(U,Z)$ and $\mathcal{B}_{w}^{C_2}(U,Z)$ are the CWEFs of the first and second constituent code, respectively. Note that only the parity check sequences from the second encoder are transmitted, hence we set $U=1$ for $\mathcal{B}_{w}^{C_2}(U,Z)$. Ultimately, the WEF of the turbo code is obtained from its CWEF by using \eqref{eq:30}, and the average BER of turbo-coded IM-FAs is given by \eqref{BER_TCM}.

\bibliography{References}
\end{document}